\documentclass[submission, Phys]{SciPost}

\begin{document}

\begin{center}{\Large \textbf{Energy and mass dependencies for the characteristics of $p_T$ regions observed at LHC energies.
}}\end{center}

\begin{center}
Mais Suleymanov
\end{center}

\begin{center}
{\bf } Baku State University \\Z. Khalilov 23, Baku Azerbijan
\\
* mais.suleymanov@bsu.edu.az
\end{center}

\begin{center}
\today
\end{center}


\section*{Abstract}
{\bf
The $p_T$ distributions of the $K^0$- and $\phi$ - mesons produced in the $pp$ collisions at $\sqrt{s}=2.76$\;$TeV$
have been analyzed by fitting them using the exponential function. It was observed that the  distributions contain several  $p_T$
regions similar to the cases with the charged particles, $\pi^0$- and $\eta$- mesons produced in the same events.
These regions could be characterized using three variables: the length of the region $L^{c}_K $ and free fitting parameters $a^{c}_K $
and $b^{c}_K $.  It was observed that the values of the parameters as a function of energy grouped around certain lines and there
are jump-like changes. These observations together with the effect of existing the several $p_T$  regions can say on discrete
energy dependencies for the $L^{c}_K $  , $a^{c}_K $ and $b^{c}_K $.  The lengths of the regions increase with the mass
of the particles. This increase gets stronger with energy. The mass dependencies of the parameters $a^{c}_K $ and $b^{c}_K $
show a regime change at a mass  $\simeq  500 MeV/c^2$.  According to the phenomenology of string theory, these results
could be explained by two processes occurring simultaneously: string hadronization and string breaking. In the experiment we
can only measure the spectrum of the hadronized particles, since we cannot access the spectrum of the strings themselves.
The string breaking effect could be a signal of string formations and the reason behind the observation of several $p_T$
regions and the jump-like changes for the characteristics of the regions.
}


\section{Introduction}
\label{sec:intro}
In the paper~\cite{1} authors argue that  $p_T$  distribution data from the LHC on invariant differential yield of the charged particles produced in $pp$ collisions at $\sqrt{s}=0.9$\:$TeV, 2.76$\:$TeV, 7$\:$TeV$ and in $Pb-Pb$ collisions at $\sqrt{s_{NN}}=2.76$\:$TeV$ with six centrality bins contained several $p_T$   regions with special properties. Similarly, the results presented in the paper~\cite{2} show the inclusive $p_T$  spectra of the $\pi^0$- and $\eta$ - mesons produced in the $pp$ collisions at LHC energies were composed of several $p_T$  regions \footnote{ The result was obtained by fitting the distributions, taking into account the errors in both axes (it was used the ROOT soft, version 5.34/02, 21st September 2012) how they are presented on the sites of hepdata. The $p_T$ distributions  were fitted  by the exponential  function ($y=a^{c}_K e^{{ b^{c}_K }p_T}$, $a^{c}_K$ and $b^{c}_K$ are free fitting parameters,upper indexes c (see Table I) show the type of  event  to which the region applies  and lower ones $K$ ($K=I$ for the first region, $K=II$ for the second one etc.)   indicate  the number of the region) in the intervals of $p_T$ between: $p^{min}_T\div p^{max}_T$, here the $p^{min}_T$    and $p^{max}_T$  are the minimum and the maximum values of $p_T$  , obtained as a result of variation of $p_T$ values to get the best fitting results. The values of $p_T$ for the boundaries of the regions for fitting $J^c_K$ are marked for  $I$ and $II$ regions  as $J^c_{I-II}$  , for $II$ and $III$ ones  as $J^c_{II-III}$ and for $ III$ and $ IV$  regions  as $J^c_{III-IV}$. The values of the $J^c_K$ were used to calculate the lengths of the regions $L^{c}_K $ as $L^{c}_I=J^c_I$; $L^{c}_{II}= J^c_{II-III}- J^c_{I-II} $ and $L^{C}_{III}= J^c_{III-VI}- J^c_{II-III} $.}. The regions could be characterized by the length $L^{c}_K $ (see footnote 1) and two free fitting parameters $a^{c}_K $  and $b^{c}_K $ (upper indexes $c$ show the type of events to which the region applies (see Table 1)  and lower ones $K$ indicate  the number of the region, see footnote 1).  It was observed that the values of the $L^{c}_K $   increase with $p_T$. The regions can be classified into two groups depending on the values of the  $L^{c}_K $,$a^{c}_K $  and $b^{c}_K $.  The values of the $L^{c}_K$ and $b^{c}_K$ for the first group don’t depend on the collision energy and the type of the particles, even though the values of $a_K^c$ increase linearly with energy, whereas the characteristics in the second group of regions show strong $s$-dependencies. It was found~\cite{2} that the ratio of the lengths (in case of the II regions) for the $\eta$-mesons ($ <L_{\eta}>$) to one for the $\pi^0$- mesons ($ <L_{\pi^0}>$) produced in the $pp$ collisions at 8 TeV is approximately equal to the ratio of their masses ($m_{\eta}$ and $m_{\pi^0}$ respectively): $ <L_{\eta}>:<L_{\pi^0}>\simeq m_{\eta}: m_{\pi^0}$. Assuming that the values of the $L^{c}_K $ are directly proportional to the string tension, the result could be considered as a smoking gun for parton/string fragmentation dynamics. We propose this explanation since in string theory the masses of elementary particles and their energies are defined by the intensity of string vibration and strangeness of the string stretch, which are depend by the tension of strings. The increase in the lengths for the $\eta$ -mesons’ regions is accompanied by an increase of the values for the parameter $b^{c}_K $.  Considering that $Q^2\simeq 1/( b^{c}_K)^2$   they calculated the values of the $\alpha_S$  for the $p_T$ regions and got that the values of $\alpha_S$ decreased with $p_T$~\cite{1}-\cite{2}. These studies have concluded that the $\eta$-mesons were produced at smaller values of $\alpha_S$ compared with

\begin{table}[htbp]
\caption{}
\centering
\begin{tabular}{|p{0.5in}|p{0.2in}|p{0.25in}|p{0.15in}|p{0.15in}|} \hline

\tiny{energy (TeV)}$\rightarrow$ \newline \tiny{particles} $\downarrow$ & 0.9 \newline  & 2.76\newline  & 7        & 8 \\ \hline
 \tiny{charged particles}                                                    &\tiny{3}       &\tiny{1}       & \tiny{4} &\tiny{ 5} \\ \hline
 \tiny{$\pi^0$-meson}                                                   &\tiny{31}      & \tiny{11}     & \tiny{41}& \tiny{51} \\ \hline
 \tiny{$\eta$-meson}                                                   &\tiny{32}      & \tiny{12}     & \tiny{42} &\tiny{ 52}  \\ \hline
 \tiny{$K^0$-meson}                                                   &\tiny{-}         & \tiny{13}   & \tiny{-} & \tiny{-}  \\ \hline
 \tiny{$\phi$-meson}                                                    &\tiny{-}       &\tiny{14}     & \tiny{-} & \tiny{-} \\ \hline

\end{tabular}
\end{table}

\noindent  that for $\pi^0$-mesons. The results show that for the first group of regions the lengths of the regions are 3-5 times greater than the lengths of neighboring lower $p_T$ regions.  For the second group of regions the lengths of the regions are 1-2 times greater than the lengths of neighboring lower $p_T$ region.  In the framework of the string fragmentation and hadronization dynamics this could mean that the particles in the group $I$ are produced from the previous generations of strings decaying into $\sim$ 3-5 new strings, while those in group II originate from the previous generations of strings decaying into $\sim$ 2 strings. They concluded that the regions might reflect features of fragmentation and hadronization of partons through the string dynamics.

In this paper we continue to analyze the $p_T$ distributions of the particles produced in the $pp$ collisions at LHC energies using the technique applied in the Revs.~\cite{1}-\cite{2}. We have analyzed  the data presented in the paper~\cite{3}  on  $p_T$ distributions of $K^0$ - and $\phi$ - mesons produced in the $pp$ collisions at $\sqrt{s}=2.76$ TeV with the goal of acquiring additional information about the energy and mass dependencies of the characteristics  of the $p_T$ regions: $L^c_K$ , $a^c_K$ and $b^c_K$ (for the $K^0$ - and $\phi$ - mesons produced in $pp$ collisions at 2.76 TeV  the index $c$=13 and 14, see Table 1).

\section{Results}
Fig. 1 shows the invariant cross sections for the charged particles, $\pi^0$- ,$\eta$-, $K^0$ - and $\phi$  - mesons (index
$c$=1, 11, 12, 13 and 14 respectively, see Table 1) production in pp collisions at 2.76 TeV~\cite{3}-\cite{5}. Actually, this is same figure as the figure 4 in the paper~\cite{2} with the addition of the new data on     $p_T$ -distributions of the $K^0$– and $\phi$-meson and recalling the data using the multiplier $10^n$ (the values of $n$  are different for the different cases ($c$) in the figure). Apparently, there no any essential differences between the distributions. However, the results of fitting these distributions  indicate some existing differences.  The Table 2 shows the best fitting results for these distributions\footnote{In the Table 2 the values of the parameter $a^c_K$ (for the case of $c$=1) were multiplied to the cross section of inelastic charged particle production ($\sigma_{inel}$) in the collisions, the values of the $\sigma_{inel} =60 mb$ had been taken from Fig. 1 in the paper ~\cite{6}.}.One can see from the data in the

\begin{figure}[htb]
\centerline{%
\includegraphics[width=12.5cm]{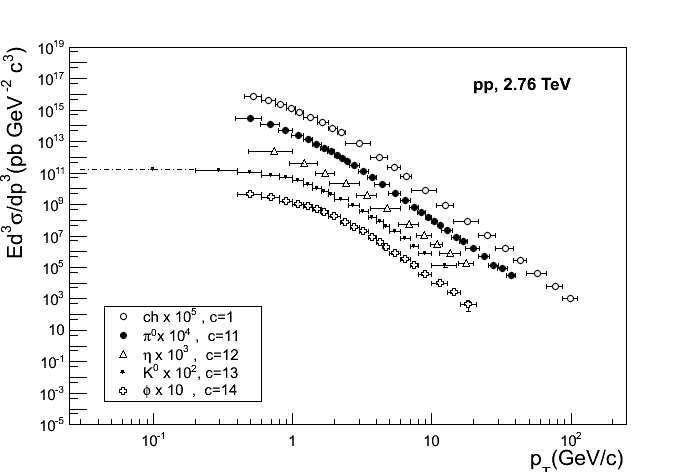}}
\caption{The invariant cross sections for the charged particles, $\pi^0$- , $\eta$-, $K^0$ - and $\phi$ - mesons (index  $c$=1;11;12;13;14 respectively, see Table 1) production in $pp$ collisions at 2.76 TeV~\cite{1}-\cite{3}.}
\label{Fig:F1}
\end{figure}

\noindent   table that $p_T$ distributions of invariant differential yield of the $K^0$ - and $\phi$ - mesons produced in the pp collisions at $\sqrt{s}$=2.76 TeV contained as well as the charged particles, the $\pi^0$- and $\eta$- mesons produced in the same events several $p_T$ regions with different values of $L^{c}_K $, $a^{c}_K $ and $b^{c}_K $.  Taking into account the results coming from the papers~\cite{1}-\cite{2} on the $p_T$ distributions of the charged particles, the $\pi^0$- and $\eta$ - mesons and the new data coming from the Table 2 for the $K^0$ - and $\phi$ - mesons we have analyzed the energy and mass dependencies of the $p_T$ regions’ parameters to reach the  aim of paper: to get an additional information on the energy and mass dependencies of the characteristics of the $p_T$ regions.

\begin{table}[htbp]
\caption{The best fit results
}
\begin{tabular}{|p{0.65in}|p{0.75in}|p{1.0in}|p{1.3575in}|p{1.55in}|} \hline
$c$\newline $\downarrow$                                                                   &                    $K$$\rightarrow$              &                 I                                       &                     II                    &                             III                                      \\ \hline
\small{$c=1$}                                                                               & $p^{min}_T\ p^{max}_T$\newline \small{$J^c_K$}\newline \small{$L^c_K$}\newline \small{{$\chi$${}^{2}$/ndf;Prob.}\newline }\small{$a^c_K$}\textit{\tiny{(pb GeV${}^{2}$c${}^{3}$})}\textit{\newline }\small{$b^c_K$} \textit{\tiny{(GeV/c)${}^{-1}$}}
& \tiny{(0.53$\pm$0.08)$\div$(3.0$\pm$0.6)}\newline
\textit{\small{J${}_{I,II}$}=(3.6$\pm$0.4)}\newline \small{(3.6$\pm$0.4)}\newline
\small{4.132/8;0.845}\newline
(3.1$\pm$0.6)10${}^{11}$
\newline\small{3.1$\pm$0.2}
& \tiny{(4.2$\pm$0.6)$\div$(18.0$\pm$3.6)}\newline
\textit{\small{J$_{II,III}$}= (21.6$\pm$2.5)}\newline
\small{(18.0$\pm$2.5)}\newline
3.458/4; 0.4842\newline
(1.8$\pm$0.8)10${}^{8}$
\newline 0.74$\pm$0.06
& \tiny{(25.2$\pm$3.6)$\div$(99.3$\pm$12.9)}\newline
 \textit{\small{J${}_{I}$${}_{II,IV}$=(100.0$\pm$12.9)}}\newline
\small{(78.4$\pm$13.3)}\newline
2.47/4;0.65\newline
(2.4$\pm$1.4) 10${}^{6}$\newline
0.13$\pm$0.01 \\ \hline
\small{$c$=11} \newline $m_{\pi^0}$=0.135 \newline GeV/$c^2$ & $p^{min}_T\ p^{max}_T$\newline $J^c_K$\newline $L^c_K$\newline \textit{$\chi$${}^{2}$/ndf;Prob.\newline}$a^c_K$\tiny{($pb GeV^{2}c^{3}$)}\small{\newline }$b^c_K$ \tiny{(GeV/c)${}^{-1}$}
& \tiny{(0.5$\pm$0.1)$\div$(3.2$\pm$0.3)}\newline
\textit{\small{J${}_{I,II}$}=(3.5$\pm$0.2)}\newline
\small{(3.5$\pm$0.2)}\newline
10.77/11;0.4631\newline
(9.5$\pm$1.9)10${}^{10}$\newline
3.0$\pm$0.1
& \tiny{(3.7$\pm$0.3)$\div$(14.9$\pm$1.1)}\newline
 \textit{\small{J$_{II,III}$=(15.9$\pm$0.8)}}\newline
\small{(12.5$\pm$0.8)}\newline
8.00318/8; 0.4332\newline
(3.9$\pm$1.4)10${}^{7}$\newline
0.78$\pm$0.04
& \tiny{(16.9$\pm$1.1)$\div$(37.3$\pm$2.7)}\newline
\textit{\small{J$_{III,IV}$=(-)}}\newline
\small{(-)}\newline \small{1.126/4;0.8901}\newline
(3.2$\pm$2.0) 10${}^{4}$\newline
0.26$\pm$0.03 \\ \hline
\small{$c$}= 12 \newline $m_{\eta}$=0.548 \newline GeV/$c^2$ & $p^{min}_T\ p^{max}_T$\newline $J^c_K$\newline $L^c_K$\newline \textit{$\chi$${}^{2}$/ndf;Prob.\newline }$a^c_K$\tiny{($pb GeV^{2}c^{3}$})\small{\newline }$b^c_K$ \tiny{(GeV/c)${}^{-1}$}
& \tiny{(0.7$\pm$0.3) $\div$(3.4$\pm$0.6)}\newline
\textit{\small{J${}_{I,II}$}= (4.1$\pm$0.6)}\newline
\small{(4.1$\pm$0.6)}\newline
0.9034/3;0.8246\newline
(1.1$\pm$0.8)10${}^{9}$\newline
2.4$\pm$0.4
& \tiny{(4.8$\pm$1.1)$\div$(13.7$\pm$2.3)}\newline
\textit{\small{J$_{II,III}$=(15.8$\pm$1.6)}}\newline
\small{(11.7$\pm$1.7)}\newline
0.6984/3; 0.8736\newline
(2.2$\pm$2.5)10${}^{7}$\newline
0.79$\pm$0.11
& \tiny{(17.8$\pm$2.2)$\div$()}\newline
\textit{\small{J$_{III,IV}$=-}}\newline
\small{-}\newline
-\newline
-\newline - \\ \hline
\small{$c$ =13} \newline $m_{K^0}$=0.498 \newline GeV/$c^2$ & $p^{min}_T\ p^{max}_T$\newline $J^c_K$\newline $L^c_K$\newline \textit{$\chi$${}^{2}$/ndf;Prob.\newline }$a^c_K$\tiny{($pb GeV^{2}c^{3}$})\small{\newline }$b^c_K$ \tiny{(GeV/c)${}^{-1}$}
& \tiny{(0.0$\pm$0.1) $\div$(4.75$\pm$0.25)}\newline
\textit{\small{J${}_{I,II}$=(5.1$\pm$0.3)}}\newline
small{(5.1$\pm$0.3)}\newline
4.132/8;0.845\newline
(3.1$\pm$0.6)10${}^{11}$\newline
3.1$\pm$0.2
&\tiny{(4.2$\pm$0.6)$\div$(18.0$\pm$3.6)}\newline
\textit{\small{J$_{II-III}$=(21.6$\pm$2.5)}}\newline
\small{(18.0$\pm$2.5)}\newline
11.12/14; 0.5973\newline
(2.6$\pm$0.2)10${}^{9}$\newline
0.1.97$\pm$0.05
& \tiny{(5.5$\pm$0.5)$\div$(12.5.3$\pm$ 2.5)}\newline
\textit{\small{J$_{III,IV}$= -}} \newline
 \small{-} \newline
1.282/3;0.7334 \newline
(1.4$\pm$1.1)10${}^{7}$\newline
 0.8$\pm$0.1  \\ \hline
\small{$c$}=14 \newline $m_{\phi}$=1.020 \newline GeV/$c^2$& $p^{min}_T\ p^{max}_T$\newline $J^c_K$\newline $L^c_K$\newline \textit{$\chi$${}^{2}$/ndf;Prob.\newline }$a^c_K$\tiny{($pb GeV^{2}c^{3}$})\small{\newline }$b^c_K$ \tiny{(GeV/c)${}^{-1}$}
& \tiny{(0.5$\pm$0.1) $\div$(5.5$\pm$0.5)}\newline
\textit{\small{J${}_{I,II}$=(6.0$\pm$0.4)}}\newline
\small{(6.0$\pm$0.4)}\newline
7.195/13;0.8918\newline
(8.4$\pm$0.8)10${}^{8}$\newline
1.82$\pm$0.05
& \tiny{(6.5$\pm$0.5)$\div$(18.5$\pm$2.5)}\newline
\textit{\small{J$_{II-III}$=-}} \newline
 \small{-} \newline
1.045/4; 0.9029\newline
(1.3$\pm$0.8)10${}^{6}$\newline
0.59$\pm$0.07
& \tiny{(17.8$\pm$2.2)$\div$()}\newline
\textit{\small{J$_{III,IV}$=-}}\newline
\small{-}\newline
-\newline
-\newline
 - \\ \hline
\end{tabular}
\end{table}

\begin{figure}[htb]
\centerline{%
\includegraphics[width=12.5cm]{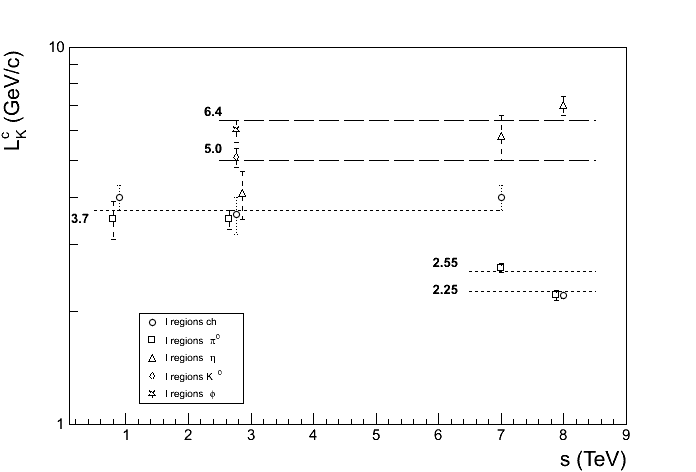}}
\caption{The energy dependences of the values of $L^c_K$ ( $I$  $p_T$ regions’ length) for the charged particles, $\pi^0$ -, $\eta$ -, $K^0$- and $\phi$-mesons produced in the $pp$ collisions at the LHC energies}
\label{Fig:F2}
\end{figure}

The Fig. 2 shows the energy dependences of the values of $L^c_K$ ($I$ $p_T$ regions’ length) for the charged particles, $\pi^0$ -, $\eta$ -, $K^0$- and $\phi$-mesons produced in the $pp$ collisions. The lines in the figure have been drawn  manually. One can see that all points relevant to values of the $L^c_I$  lie on 5 lines at $L^c_I\simeq $2.25;2.55;3.7;5.0 ; 6.4 GeV/c.  The figure also shows that for the charged particles, the energy dependence appear jump-like at $\sqrt{s}$ = 8 TeV whereas the values of the $L^5_I$   decrease from the values of 3.7 GeV/c and occur on line at the level  2.25 GeV/c. For the values of $L^c_I$   in case of  $\pi^0$-mesons the $s$-dependence starts to observe at $\sqrt{s}$ =7 TeV. The corresponding points for the values of $L^c_I$ at 7 and 8 TeV move from the line at level 3.7 GeV/c to one at levels 2.55  GeV/c and 2.25 GeV/c again jump-like. The values of $L^c_I$  for the $\eta$-mesons are on the line at level 3.7 GeV/c for the  $\sqrt{s}$ = 2.76 TeV. With increasing the energy in the interval of  $\sqrt{s}$ = 7-8 TeV the values of the $L^c_I$  for the eta mesons increase jump-like and occur on the line at the level 5.0 GeV/c and 6.4 GeV/c. The values of the $L^{13}_I$  for $K^0$ -mesons and $L^{14}_I$ for $\phi$-mesons are on the line at level 5.0 GeV/c and 6.4 GeV/c too. One can conclude that with energy the values of $L^c_I$, in the interval of: $\sqrt{s}\leq$2.76 TeV don’t depend on energy for  the charged particles, $\pi^0$-, $\eta$-mesons; as soon as the energy reaches the values of $\sqrt{s}$ =7 TeV the values of $L^c_I$  for all charged particles and $\pi^0$-mesons decrease though for the $\eta$ - mesons the values of lengths increase  jump-like.  If we take into account that the mass of the $\eta$-,$K^0$- and $\phi$- mesons much more  than the average mass of all charged particles (produced in pp collisions) and $\pi^0$-mesons,  so  it can then be concluded that  the $p_T$  regions’  lengths change jump-like in the interval of energy $\sqrt{s}\geq$ 7 TeV whereas the lengths of light particles decrease , but the ones for the heavy particles increase. The junk-like changes could mean that in the first region lengths might change discretely with energy.

The energy dependencies of the  $II$  $p_T$ regions’ lengths for the charged particles, $\pi^0$- and $\eta$-mesons produced in the $pp$ collisions are shown in the Fig. 3. One can see that all defined values for the lengths are approximately on 3 lines at level $L^c_{II}\simeq$ 14.8; 4.5;2.5

\begin{figure}[htb]
\centerline{%
\includegraphics[width=12.5cm]{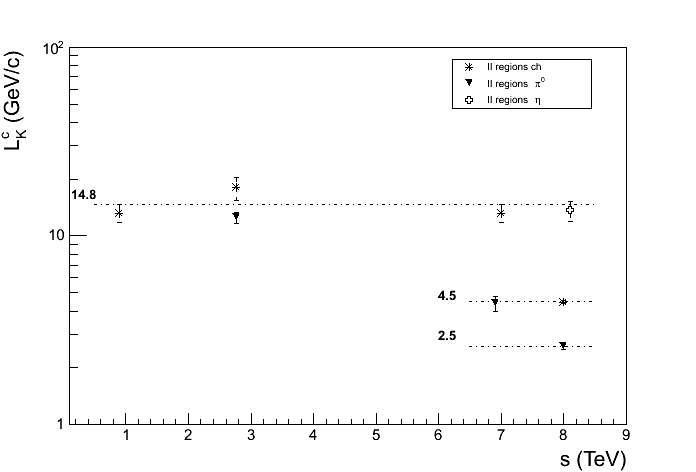}}
\caption{The energy dependencies of the values of $L^c_{II}$  ($ II$ $p_T$ regions’ length) for the charged particles, $\phi^0$- and $\eta$-mesons produced in the $pp$ collisions at the LHC energies.}
\label{Fig:F3}
\end{figure}

\noindent  GeV/c. The values of the $L^c_{II}$ for all charged particles do not depend on energy in the interval of $\sqrt{s}\leq$ 7 TeV. The energy dependence occurs at $\sqrt{s}$ = 8 TeV. Whereas, the values of length for the charged particles move from line at $L^c_I\simeq$  14.8 GeV/c to one at $L^c_{II}\simeq$ 4.5 GeV/c jump-like. In case of $\pi^0$-mesons we had 3 points for the values of lengths, for $\sqrt{s}$ = 2.76 TeV the values of $L^{11}_{II}$ are on the line at $L^c_{II}\simeq$ 14.8 GeV/c, for $\sqrt{s}$ = 7 TeV the values of $L^{41}_{II}$ are on the line $L^c_{II}\simeq$4.5 GeV/c and for $\sqrt{s}$ = 8 TeV the values of $L^{51}_{II}$  do on the line  $L^c_{II} \simeq$ 2.5 GeV/c. The single values of length for the eta mesons at $\sqrt{s}$ = 8 TeV ($L^{52}_{II}$) is on the line $L^c_{II}\simeq$ 14.8 GeV/c. It means that the $L^{52}_{II}> L^{51}_{II}$  which means that the lengths of the $II$  $p_T$ regions for the eta mesons are greater than for the neutral pions. Again it can be noted that there is jump-like energy dependence of the lengths for $II$ $p_T$ regions - discrete energy dependence.

\begin{figure}[htb]
\centerline{%
\includegraphics[width=12.5cm]{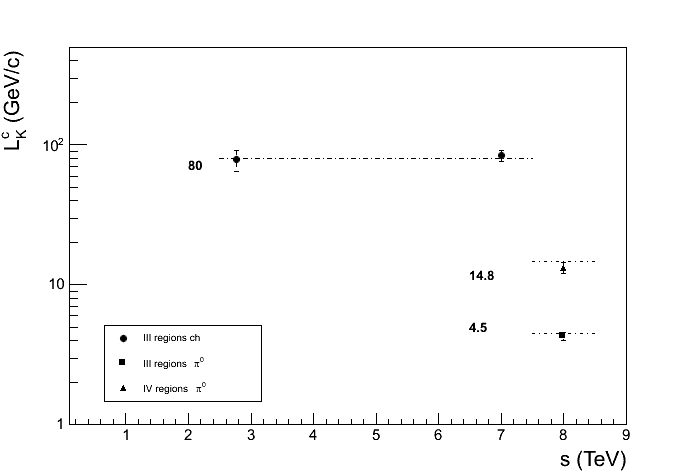}}
\caption{The energy dependences of the values of $L^c_K$ of the $ III-IV$ $p_T$ regions’  lengths for the charged particles, $\pi^0$- and $\eta$-mesons produced in the $pp$ collisions at the  LHC energies.}
\label{Fig:F4}
\end{figure}

The Fig. 4 shows the energy dependences of the values of the lengths for the $III$ and $IV$ $p_T $ regions for the charged particles and $\pi^0$-mesons produced in the $pp$ collisions. The lines in the figure have been drawn manually.  These show that the all values of the $L^c_{III-IV}$ roughly lie down on 3 lines at $L^c_{III}\simeq$ 80;14.8  and 4.5 GeV/c.  It is interesting that the line at $L^{51}_{III}\simeq$14.8 GeV/c has been observed for the $II$ $p_T$ regions (see previous figure)  and the lines at $L^{51}_{IV}\simeq$4.5 GeV/c has been observed for the $I$ $p_T$ regions (see Fig. 2). The last again could be considered as a signal on discrete changes with energy for the lengths of the $ p_T$  regions.

So it can be concluded that with energy the lengths of the regions change jump-like, whereas the lengths for the light mesons decrease and   these are increased for the heavy mesons.  The last result is seen more cleanly from the Fig. 5,  which shows the mass dependence for the values of $L^c_K$  (in cases 2 or more points in the distributions). The lines have been drawn  manually. One can see that with mass (in the $I$  $p_T$ region at $\sqrt{s}$ = 2.76 TeV)  the values of the lengths  increase slowly and in the $II$  $p_T$  regions at $\sqrt{s}$  =2.76 TeV (we have had only 2 points)   the values of the lengths show an independence on mass.  At $\sqrt{s}$  = 7 TeV the values of  $L^c_K$  increase sharper than at 2.76 TeV ,  gets more strong at energy 8 TeV and the relation holds:$ <L_{\eta}>:<L_{\pi^0}>\simeq m_{\eta}: m_{\pi^0}$ (which was observed in the paper~\cite{2}).
The dependencies of the lengths of $p_T$  regions on the mass of particles and the strengthening of this dependence with increasing energy of colliding protons and the result on $ <L_{\eta}>:<L_{\pi^0}>\simeq m_{\eta}: m_{\pi^0}$ together with jump-like energy dependencies can be arguments in favor of string theory.

\begin{figure}[htb]
\centerline{%
\includegraphics[width=12.5cm]{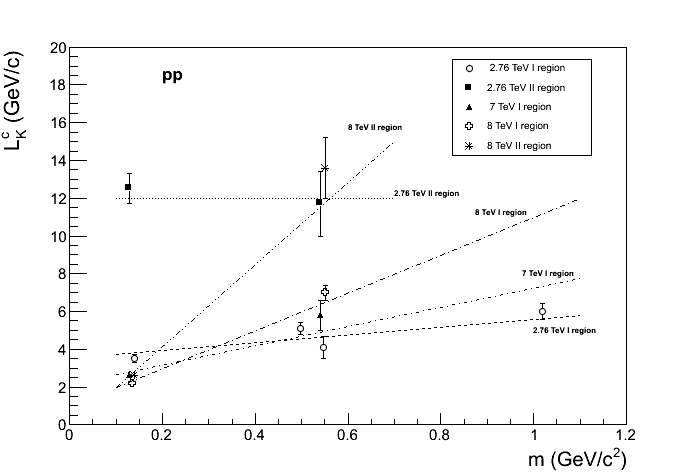}}
\caption{The mass dependences for the regions’ lengths.}
\label{Fig:F5}
\end{figure}

The energy dependencies for the parameters $a^c_I$ ($I$  $p_T$ regions) are shown in the Fig. 6. It can be seen from the figure\footnote{The parameter  $a^c_K$  has unit $pb\ GeV^{-2}c^3$}  that the values of $a^c_I$ for charged particles and $\pi^0$- mesons increase with energy.  We tried to fit these distributions by the function: $y=gs^n$ ( where $g$ and $n$ are free fitting parameters)\footnote{ We could not get satisfactory results for fitting (due to the small number of fit points). Therefore, we fixed the values of $g$ and found the values of $n$, then we fixed the found values of $n$ and refined the values of $g$ . Thus, we chose the values $g$ and $n$.
}. The results of the fitting are shown in Fig. 6  (and in Table 3). It turned out that the distributions for charged particles fitted well by functions of: $y=2.1 10^{11}s^{0.36\pm0.05}$ , and the data for $\pi^0$-mesons are in satisfactory agreement with the function : $y=4.3 10^{10}s^{0.89\pm0.06}$  only at energies of 0.9–7.0 TeV. At the energy 8.0 TeV, an increase in $a^c_K$     is observed and the corresponding point moves and how would it “jumps” on the line $y=2.1 10^{11}s^{0.36\pm0.05}$. We see again like in case of lengths the jump-like changing for the parameter $a^c_K$ in the $I$  $p_T$ region. For the values of $a^c_I$    in the case of  $\eta$-mesons from region $I$, we had 3 points (at energies 2.76, 7 and 8 TeV) only. The result of fitting these data by a function: $y=4.2 10^{8}s^{1.5\pm0.2}$ (only for two points corresponding to energies 2.76 and 7.0 TeV) is shown in the figure by a dashed line ( in the Table 3). As the energy increases to 8 TeV, the values of $a^{52}_I$   decrease and do not lie on the line: $y=4.2 10^{8}s^{1.5\pm0.2}$. The point occurs on the line describing by the function $y=2.9 10^{5}s^{4.80\pm0.09}$  (which describes well the energy dependence of the parameter $a^c_K$ for  the neutral pions from $II$  $p_T$  regions , see Fig. 7). It is interesting that the single points for the values of $a^c_K$  for $K^0$ - and $\phi$- mesons (at 2.76 TeV) are on the line: $y=4.2 10^{8}s^{1.5\pm0.2}$ too.

\begin{figure}[htb]
\centerline{%
\includegraphics[width=12.5cm]{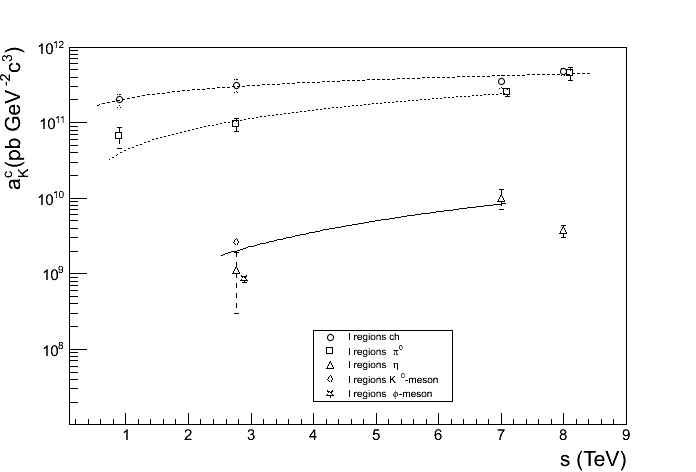}}
\caption{The energy dependences for the parameters $a^c_I$ ($I$  $p_T$ regions).}
\label{Fig:F6}
\end{figure}

Thus one can say that the values of the parameters $a^c_K$ as a function of energy are grouped around 4 lines with a power-law dependence on energy and show jump-like behavior at high energy as a signature to discrete changing.
 The data on $a^c_K$    for charged particles from the $II$ region of $p_T$  (Fig. 7) were well fitted by a function: $y=8.2 10^{7}s^{(0.8\pm 0.1)}$  in the energy range 0.9–7.0 TeV(the fitting results are shown in the Fig. 7), but  the data for energies of 8 TeV strongly deviate from this dependence, this point is close to the line: $y=2.9 10^{5}s^{(4.80\pm0.09)}$. For neutral pins from this region, a satisfactory fit was obtained using a function: $y=4.3 10^{10}s^{(5.36\pm0.09)}$. Here one can see a very strong dependence of $a^c_K$ on energy.
 In the case of $\eta$-mesons from the $II$ $p_T$ region, we have only one point for the value  $a^c_K$    at an energy of 8 TeV (it is approximately on the solid line at the level of 6.5 $10^6$). For the values of $a^c_K$  of $K^0$ - and $\phi$- mesons from the second $p_T$  region, we also had one point at  energy of 2.76 TeV and we can write that $a^{11}_{II}\simeq a^{13}_{II}\simeq a^{14}_{II}$ (we recall that here  the  $a^{11}_{II}$, $a^{13}_{II}$, $a^{14}_{II}$  are the values of the free fitting parameter  $a^c_K$ at an energy of 2.76 TeV for neutral pions, $K^0$ - and $\phi$- mesas, respectively, from the second $p_T$ region).

 For the  $3^{rd}$  $p_T$  regions (see Fig.7) there have not had even  3 defined points for the values of $a^c_K$, and only in the case of charged particles the values of the $a^c_K$ were determined at two energies: 2.76 and 7 TeV. It can be seen that the values of $a^c_K$ decrease with energy (the distribution was fitted by the function: $y=2.57 10^8s^{-(4.6\pm0.2)}$. Let us recall that the lengths of the regions in this case did not change.For the case of $\pi^0$-mesons from the $III$ $p_T$ region we have had one point only
 (at energy 8 TeV, which is around the level $a_K^c\simeq 6.5\:10^6$).
 There is a single point for the $IV$ $p_T$ regions’ pions (it is around the values of 
 $a_K^c\simeq 6.5\:10^6$ ,see Fig.7).

\begin{figure}[htb]
\centerline{%
\includegraphics[width=12.5cm]{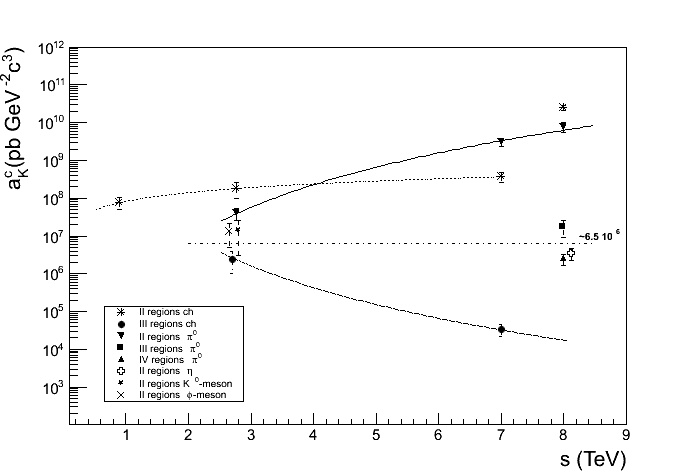}}
\caption{The data on $a^c_K$ for the particles from the $II-IV$ regions of $p_T$.}
\label{Fig:F7}
\end{figure}

\begin{table}[htbp]
\caption{}
\begin{center}
\begin{tabular}{|p{22pt}|l|l|l|l|}
\hline
&
&
I&
II&
III \\
\hline
&
$y$&
$y={2.1\, 10}^{11}s^{({0.36\pm0.05})}$&
 $y={8.2\, 10}^{7}s^{({0.8\pm0.1})}$&
 $y={2.6\, 10}^{8}s^{-({4.6\pm0.2})}$ \\
\textit{ch}&
$s$&
\textit{0.9 -- 8 TeV}&
\textit{0.9 -- 7 TeV}&
\textit{2.76 -- 7 TeV} \\
&
$\chi {}^{2}/ndf ;  Prob.$&
1.372/3 ; 0.712&
0.000684/2;0.9996&
0.03012/1; 0.8622 \\
\hline
&
$y$&
$y={4.3\, 10}^{10}s^{({0.89\pm0.06})}$&
$y={2.9\, 10}^{5}s^{({4.8\pm0.1})}$&
- \\
$\pi^{0}$&
$s$&
\textit{0.9-7.0 TeV}&
\textit{2.76 -- 8 TeV}&
- \\
&
$\chi {}^{2}$/ndf ;  Prob.&
1.991/2; 0.3695&
0.6158/2 ; 0.735&
- \\
\hline
&
$y$&$y={4.2\, 10}^{8}s^{({1.5\pm0.2})}$
&
-  &
- \\
$\eta $&
$s$&
\textit{2.76 - 7 TeV}&
-&
- \\
&
$\chi^2/ndf ;  Prob.$&
1.563/1; 0.2112&
-&
- \\
\hline
\end{tabular}
\label{tab1}
\end{center}
\end{table}

The energy dependences for the parameters $b_K^c$ are shown in
the Fig. 8 for the I $p_{T}$ regions' data. To fit the distributions, we have
used the function  $y=gs^n$ (the results of the fitting are in the
Table 4). One can see (from the Fig.8 and Table 4) that for charged
particles from the $I$ $p_T$ region, the values of $b^c_K$ are
almost independent of $s$ and lie on the line $y=(3.00\pm0.09)s^(-(0.015\pm0.010))$ at $s\cong $0.9-8 TeV and for the $\pi
^{0}$-mesons from this region , the values of $b^c_K$ depend
on $s$ as $y=(2.9\pm0.1)s^{(0.07\pm0.03)}$ which is very close to one for the charged
particles. The data on $b_K^c$ for the $\eta $-mesons from the I
$p_{T}$ region shows the behavior which could describe by function $ y=(4.3\pm1.1)s^{-(0.4\pm0.1)}$ at $s\cong $ 2.76-8 TeV. In this regions for $K^{0}$- and
$\varphi $-mesons we have had only one point which are at the intersection
of two lines defined by functions :
$ y=(3.00\pm0.09)s^{-(0.015\pm0.010)}$ and  $y=(2.9\pm0.1) s^{(0.07\pm0.03) }$ .

The Fig . 9 shows the energy dependencies of the $b_K^c $ for the
particles from the $II,III$ and $IV$ $p_{T}$ regions. One can see that for the
charged particles, the values of $b_K^c$ depend weakly (see
Table 4 ) on $s$ and lie on the line of 
$y=(0.85\pm0.04)s^{-(0.07\pm0.04)} $ at $s\cong $ 0.9-7 TeV, an energy dependence is observed
at transition to the area of energy 8 TeV, whereas the values of the $b_K^c$ increase jump-like and is on the line
$ y\simeq(0.38\pm0.04)s^{(0.71\pm0.06)}$. This line was found for the values of
$b_K^c $ in case of neutral pions at the energies 2.76-8 TeV. The
values of $b_K^c $ for $III$ region charged particles show non
dependence on energy and lie on the $y=(0.11\pm0.02)s^{(0.1\pm0.1)}$ at the
energies 2.76-7 TeV. In the cases of : $III$ and $IV$ regions for
neutral pions; II region for eta mesons ; II regions for $K^{0}$- and
$\phi$-mesons we have had only one point.

The above results for the behaviors of parameters $b_{K}^{c}$ indicate that
in the interval of \textit{s \textgreater }2.76 TeV there appear strong energy and mass dependences
for the values of $b_{K}^{c}$ . In the paper~\cite{1}-\cite{2} it has been discussed
that the observed $p_{T}$ regions at LHC could relate to the string dynamics
of fragmentation hadrons and hadronization of partons. For the
neutral pions the values of $b_{K}^{c}$ increase with $s$ but for the eta
mesons ones decrease with energy. As it has been discussed in the paper~\cite{1}
the parameter $b_{K}^{c}$ might connect with the distance between partons in
the parton strings that is why the result can mean that eta mesons due to
strange quarks containing produced in the shorter distance then the distance
of the neutral pions production.

\begin{figure}[htb]
\centerline{%
\includegraphics[width=12.5cm]{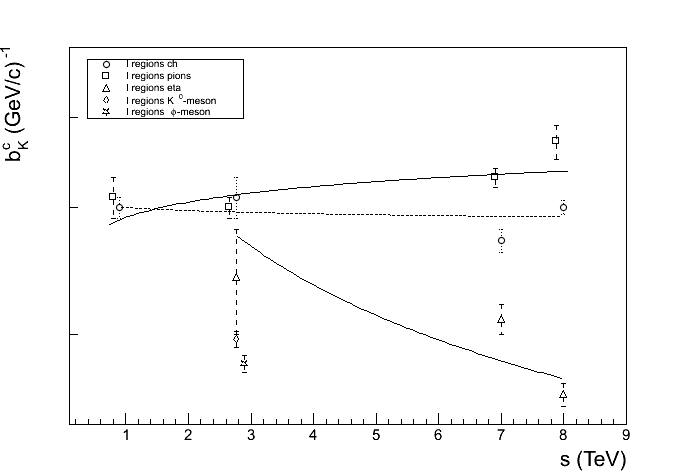}}
\caption{The energy dependences for the parameters $b^c_K$.}
\label{Fig:F8}
\end{figure}

\begin{table}[htbp]
\caption{}
\begin{center}
\begin{tabular}{|p{0.2in}|p{0.55in}|p{1.250in}|p{1.3575in}|p{1.55in}|}
\hline
&
&
I&
II&
III \\
\hline
&
$y$&
$y=(3.00\pm0.09)s^{(0.015\pm0.010)}$&
 $y=(0.85\pm0.04)s^{-(0.07\pm0.04)}$&
$y=(0.11\pm0.02)s^{(0.1\pm0.1)}$ \\
\textit{ch}&
$s$&
\textit{0.9 -- 8 TeV}&
\textit{0.9 -- 7 TeV}&
\textit{2.76 -- 7 TeV} \\
&
\tiny{$\chi^{2}$/ndf;Prob.}&
\tiny{6.828/3  ;  0.07758}&
\tiny{1.011/1;0.3147}&
- \\
\hline
&
$y$&
 $y=(2.9\pm0.1)s^{(0.07\pm0.03)}$&
 $y=(0.38\pm0.04)s^{-(0.71\pm0.06)}$&
- \\
$\pi^{0}$&
$s$&
\textit{0.9-7.0 TeV}&
\textit{2.76 -- 8 TeV}&
- \\
&
\tiny{$\chi^{2}$/ndf;Prob.}&
\tiny{5.662/2; 0.05895}&
\tiny{0.7199/1; 0.3962}&
- \\
\hline
&
$y$&
$y=(4.3\pm1.1)s^{-(0.4\pm0.1)}$&
-&
- \\
$\eta $&
$s$&
\textit{2.76 - 8 TeV}&
-&
- \\
&
\tiny{$\chi^2$/ndf;Prob.}&
\tiny{9.634/1; 0.00192}&
-&
- \\
\hline
\end{tabular}
\label{tab1}
\end{center}
\end{table}
\begin{figure}[htb]
\centerline{%
\includegraphics[width=12.5cm]{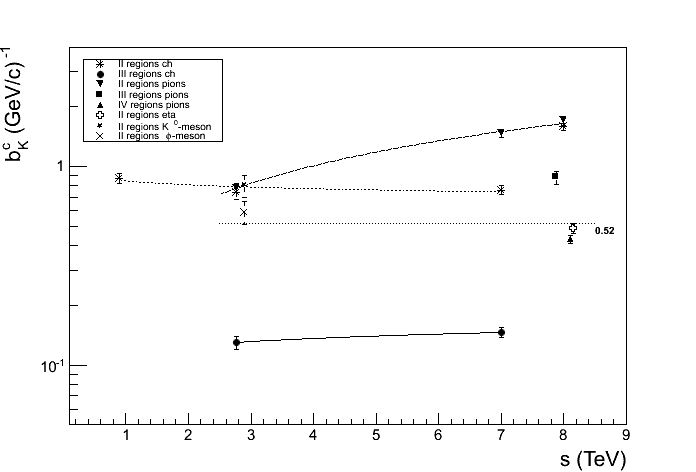}}
\caption{The energy dependences for the parameters $b^c_K$  ($II-IV$   $p_T$ regions).}
\label{Fig:F9}
\end{figure}

Fig. 10 shows the energy dependencies, of the lengths $L_{K\, }^{c}$
multiplied by the values of the free fitting parameter $b_{K}^{c}$ in the
same regions ($L_{K\, }^{c}\ast b_{K}^{c})$ for the charged particles,
$\pi^{0}$- , $\eta $-, $K^{0}$- and $\phi $ - mesons produced in \textit{pp}
collisions at LHC energies\footnote{ The data for the charged particles were
taken from the paper~\cite{1} and the ones for the $\pi^{0}$- and $\eta - $mesons
were done from the paper [2]}. It can be seen that with energy, the values
of $L_{K\, }^{c}\ast b_{K}^{c}$ for most of cases remain unchanged at the
level of $L_{K\, }^{c}\ast b_{K}^{c}\cong 11$. The deviations from this
value begin to appear at energy of 7 TeV for charged particles and neutral
pions from $II$ $p_{T}$ region. For these events the values of $L_{K\,
}^{c}\ast b_{K}^{c}\cong 8$ or 6 . At energy 8 TeV, deviations get stronger
and have been observed for almost all cases (except for one case with c$=$
52 - for the eta mesons at 8 TeV). Now the corresponding points are on the
lines at 8, 6 and 4. Again one can say that with energy the values of the
$L_{K\, }^{c}\ast b_{K}^{c}$ change jump-like as a signal on discrete change
the values of $L_K ^c*b_K^c$ .

\begin{figure}[htb]
\centerline{%
\includegraphics[width=12.5cm]{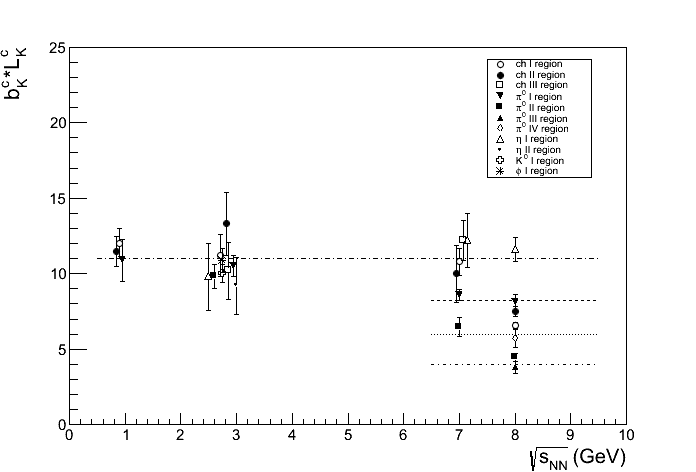}}
\caption{The energy dependencies, of the lengths $L_{K }^c$ multiplied by the values
of the free fitting parameter $b_K^c$.}
\label{Fig:F10}
\end{figure}

Fig. 11-12 show the mass dependence of the values of $a_{K}^{c}$ and
$b_{K}^{c}$ for the cases when there are at least 2 defined points for the
values of $a_{K}^{c}$ and $b_{K}^{c}$ . One can see that with mass in the $I$
$p_{T}$ region at $\surd s \quad =$ 2.76 TeV the values of the parameters
$a_{K}^{c}$ and ${\, b}_{K}^{c}$ first decrease in the interval of mass m
\textless 500 MeV/c$^{2}$ and then are almost independent of mass in the
interval of m \textgreater 500 MeV/c$^{2\, }$. But for the $II$ $p_{T}$ regions
at $\surd $s$=$2.76 the values of the parameters $a_{K}^{c}$ and $b_K^c$ don't depend on mass. At $\surd s=$7 TeV the values the
parameters $a_K^c $ and $b_K^c $ decrease. At a
further increase the energy at 8 TeV the values of the $a_K^c$
and $b_K^c $ decrease sharply. So one can note that unlike
parameter $L^c_K$ the mass dependencies of the parameters $a_K^c $ and $b_K^c$ show regime change at mass $m \quad \cong $
500 MeV/c$^{2\, \, }$.

\begin{figure}[htb]
\centerline{%
\includegraphics[width=12.5cm]{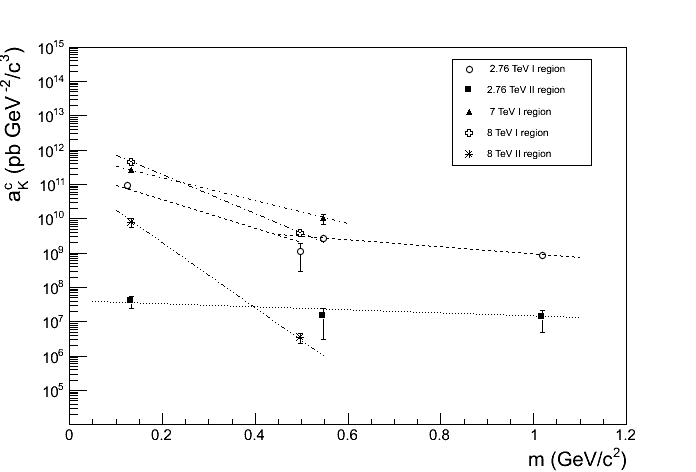}}
\caption{The mass dependence of the values of $a_K^c$. }
\label{Fig:F11}
\end{figure}

\begin{figure}[htb]
\centerline{%
\includegraphics[width=12.5cm]{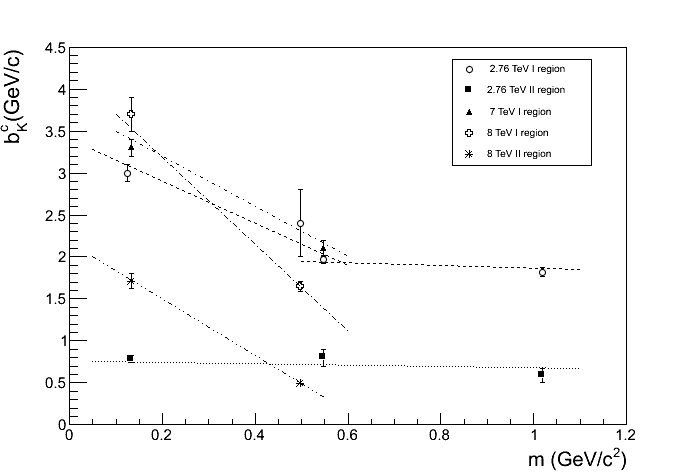}}
\caption{The mass dependence of the values of $b_K^c$.}
\label{Fig:F12}
\end{figure}

Before the conclusion, we wrote up the list of the main results:

\begin{enumerate}
\item The $p_{T}$ distribution data on the invariant differential yield of the $K^{0}$- and $\phi - $mesons produced in the pp collisions at $\sqrt s =$2.76 TeV contained several $p_{T}$ regions which could be characterized by the lengths of the regions and the values of the free fitting parameters $a_{K}^{c}$ and $b_{K}^{c}\, $.
\item All of the points relevant to values of the lengths for the I $p_{T}$ regions as a function of energy lie down on five lines at $L_{I}^{c}\cong 2.25,\, 2.55,\, 3.7,\, 5.0$, 6.4 GeV/c; the II $p_{T}$ regions' points are approximately on three lines at levels $L_{II}^{c}\cong $ 14.8, 4.5, 2.5 GeV/c; the III and IV $p_{T}$ regions' points are  roughly lie on three lines at $L_{III-IV}^c\simeq$80,14.8 and 4.5 GeV/c. The energy dependence shows jump-like changes and indicating discrete changes with energy.
\item The lengths of the regions increase with their masses. The mass dependence gets stronger with energy (at 8 TeV the following relation holds:$<L_{\eta}>:<L_{\pi^9}>\simeq m_{\eta}:m_{\pi^0} $ which was obtained in the paper in~
    \cite{2}).
\item The values $a^c_I$ and $a^c_{II}$ as a function of energy are grouped around several lines with a power-law dependence on energy and show jump-like behavior at high energies as a signature of discrete changes.
\item The resulting behavior of the parameter $b_K^c$ tells us that in the interval of \textit{s \textgreater }2.76 TeV there is a strong energy dependence. For the neutral pions the values of $b_K^c $ increase with $s,$ but for the eta mesons $b_K^c $ decreases with energy. As the energy increases the values of $L^c_K*b^c_K$  for most of the cases remain unchanged at the level of $L^c_K*b^c_K\simeq11,8,6,4$, displaying jump-like changes as a signal of a discrete behavior.
\item The mass dependences of the parameters $a_K^c$ and $b_K^c$ show regime change at 500 MeV/c$^{2\, \, }$.
\end{enumerate}

\section{Conclusion}
The $p_{T}$ distribution data on invariant differential yield of the
$K^{0}$- and $\phi - $mesons produced in the pp collisions at $\sqrt s
=$2.76 TeV contain several $p_{T}$ regions, which have been observed for
the charged particles, neutral pions and eta mesons. The regions could be
characterized by their lengths $L_{K}^{c}$ and the free fitting parameters
$a_{K}^{c}$ and $b_{K}^{c}$. The values of $L_{K}^{c}$, $a^c_K$
and $b^c_K$ as a function of energy grouped around several lines
and show jump-like changes behavior with energy. These observations together
with the effect of existing the several $p_{T}$ regions can say on discrete
energy dependencies for the $L_{K}^{c}$, $a^c_K$
and $b^c_K$.

The lengths of the regions increase with the mass of the particles and the
rate of increase gets stronger with energy. The mass dependencies of the
parameters $a^c_K$
and $b^c_K$ show regime change at
mass $\cong $ 500 MeV/c$^{2}$.

According to the phenomenology of string theory, the results could mean that
parton strings of the very first generation with maximum tensions $T_{max}$
formed immediately after the collision either hadronize or decay forming the
next generation of strings with tensions $T^{\mbox{'}}$ \textless T$_{max}$.
The newly formed strings can, again, either hadronize or decay and create
strings of the next generation with tensions $T {\mbox{"}}> T{\mbox{'}}$, etc., up to
the minimum tensions $T_{min}$ after which the string decay halts. That is,
two processes occur simultaneously: string hadronization and string
breaking. In the experiment we measure the spectrum of the hadronized
hadrons, since we cannot have a spectrum of the strings themselves. The
string breaking effect is the reason of observing several $p_{T}$ regions
which has a discrete nature. The latter can explain the observed results on
jump-like change of the characteristics of the regions and grouping the
values of the characteristics around certain lines of energy.

Finally, it is important to note that in the experiment, we can see only the
signature of the last generations of strings, which are in the area of our
$p_{T}$ measurements. It is very difficult to have signature for the first
generation of strings, since for this it is necessary to have measurements
of the $p_{T}$ in the interval up to several TeV / c.


\nolinenumbers

\end{document}